\documentclass[12pt]{article}
\usepackage{amssymb,amscd,amsmath,amsthm}
\usepackage{latexsym,amstext,cite}
\usepackage{color}

\def\be{\begin{equation}}
\def\ee{\end{equation}}
\def\bea{\begin{eqnarray}}
\def\eea{\end{eqnarray}}
\def\bean{\begin{eqnarray*}}
\def\eean{\end{eqnarray*}}

\def\D{{\mathbb D}}

\begin{document}


\thispagestyle{empty}
\hfill \today
\vspace{1.5cm}
\begin{center}
{\LARGE\bf Algebraic Image Processing}
\end{center}

\bigskip

\begin{center}{Enrico Celeghini}
\bigskip

{\it
Dipartimento di Fisica, Universit\`a di Firenze and INFN-Sezione di Firenze\\
 I50019 Sesto Fiorentino, Firenze, Italy.
\medskip

Departamento de F\'isica Te\'orica, At\'omica y \'Optica and IMUVA. \\
Universidad de Valladolid, 
47011 Valladolid, Spain.
}
\medskip

e-mail: celeghini@fi.infn.it.
\end{center}
\bigskip

\begin{abstract}

  We  propose  an approach to image processing 
  related to algebraic operators acting in the space of images.
  In view of the interest in the applications in optics and computer science,
  mathematical aspects of  the paper have been simplified as much as possible.
  Underlying  theory, related to rigged Hilbert spaces 
  and Lie algebras, is discussed elsewhere.   
 
 OCIS codes 100.2000, 100.2980, 110.1085, 110.1758

\end{abstract}

\section{Introduction}

The fundamental problem in image analysis is that the information contained in an image is 
immense, too much in order that the human mind can manage it.
So in science we have to disregard a large part of the contained information and isolate
the required one, quite more restricted, in function of our specific interests. 

We suggest that this work to clean the signal from forgettable elements and to
highlight the relevant information  could be improved 
by the mathematical theory of operators acting in the space of images.

Same application of these ideas, samehow similar to adaptive optics, are 
described in the following.
Yet, while in adaptive optics the perturbations to be removed are related to the 
phases of a complex function\, $f(r,\theta)$ our approach, that we could call
Algebraic Image Processing (AIP), acts on the 
module of the image\, $|f(r,\theta)|$.
So, we suggest that the image $|f(r,\theta)|^2$ registered by an 
experimental apparatus can be assumed not as the final result of the optical measure but 
possibly as an intermediate step to be further elaborated by  a computer
program based on operators  acting in the space of images.
AIP is indeed a soft procedure and operates on a set of pixels
to obtain another set of pixels 
independently from the causes of the effects we wish to eliminate.
  
Moreover AIP is a general theory of the connections of images between them and 
thus can be applied also outside the cleaning of images to any manipulation of images.

We restrict ourselves in this paper to practical aspects relevant in 
optics and computer science. 
Underlying mathematics, related to rigged Hilbert spaces and
Lie groups, is considered in detail in \cite{Ce17} and, at higher theoretical level, in
\cite{ReSi, CeGaOl16, CeGaOl17}.  We recommend to the reader interested in the theory 
to look there and references therein.

 Features of the proposed approach are: 
 \begin{enumerate}
 \vspace{-.4cm}
 \item  It can be used on-line and also off-line if, at later time, more accuracy is required.
\vspace{-.3cm}
\item It is a soft procedure, cheap and without mechanical moving parts.
\vspace{-.3cm}
\item It allows to consider together images of different origin and frequencies like optical and 
radio images.
\end{enumerate}
\vspace{-.2cm}


 The relevant operative points are:
\begin{itemize}
\vspace{-.4cm}
 \item
The vector space of images on the disk has as a basis the Zernike functions \cite{BoWo}.
\vspace{-.3cm}
\item
The space of images can be integrated with the operators in the space of images i.e. 
the operators that transform Zernike functions into Zernike functions.
\vspace{-.3cm}
\item 
This space of images and its operators define
the unitary irreducible representation
$D_{1/2}^+ \otimes D_{1/2}^+$\;\, of the Lie algebra\, $su(1,1) \oplus su(1,1)$\,.
\item
\vspace{-.3cm}
Every operator of AIP can be written as a polynomial in the 
generators of the algebra $su(1,1)\oplus su(1,1)$ 
\vspace{-.3cm}
\item
and can be computed applying this polynomial to the Zernike functions.
\end{itemize}

In sect.2 we summarize the vector space of images. In sect.3 we introduce the 
operators acting on this space, their algebraic structure and their realization. 
In sect.4 we give a description of a possible modus operandi that could be realized
in automatic or semi-automatic way by a computer program. Color  images can also be
introduced by means of an additional factorization of a color code.

\section{Vector Space of Images}

Radial Zernike polynomials $R^m_n(r)$ can be found in \cite{BoWo}
and are real polynomials defined for\, {$0\leq r \leq 1$}
such that  $R^m_n(1) = 1$, where   
{$n$}\, is a natural number and {$m$}\,an integer, such that\, 
{$0\leq m\leq n$}\, and\, {$n-m$}\, is even.

As the interest in optics is focused on real functions defined on the disk,
starting from $R_n^m(r)$ it is usual to introduce, with $0 \leq \theta < 2 \pi$, the functions
\[
{\cal Z}^{-m}_n(r,\theta) := R^m_n(r) \;{\rm sin}(m \theta)\qquad {\cal Z}^{m}_n(r,\theta) := 
R^m_n(r) \;{\rm cos}(m \theta)
\]
and, because only smooth functions are normally considered, to take into account
only low values of $n$ and $m$, summarized in a unique sequential
index \cite{No}.

However, in a general theory, functions are not necessarily smooth and formal properties 
are relevant. We thus came back to the classical form of Born and Wolf in the complex space\cite{BoWo}:
\[
Z_n^m(r,\theta) := R_n^{|m|}(r)\; e^{{\bf i} m \theta}
\]
with $n$ natural number, $m$ integer with $n-m$ even and $-n \leq m \leq n$.
The symmetry can be improved  writing $n$ and $m$ in function of 
two arbitrary independent natural numbers $k$ and $l$  \cite{Du}
\[
n = k+l\,\qquad m = k-l\qquad\qquad (k=0,1,2,...;\; l=0,1,2,...)
\]
and introducing a multiplicative factor. The Zernike functions we consider here are thus 
\begin{equation}\label{-1}
V_{k,l}(r,\theta)\; := \sqrt{k+l+1}\;\; R_{k+l}^{|k-l|}(r)\; e^{{\bf i}(k-l)\theta}
\end{equation}
and depend from two natural numbers $k$ and $l$ and two continuous variables 
$r$ and $\theta$.
The functions $V_{k,l}(r,\theta)$  are an orthonormal basis in the space\, $L^2(\D)$\, 
of square integrable complex functions defined on the unit disk $\D$ as:
\begin{align}\label{1a}
\begin{split}
\frac{1}{2\pi}&\, \int^{2 \pi}_0 d \theta \int^1_0 dr^2 \; {V_{k,l}(r,\theta)}^* \;\;    {V_{k',l'}(r,\theta)}\;\;\; =\; 
\delta_{k,k'}\, \delta_{l,l'}\\
\frac{1}{2\pi}&\; \sum_{k=0}^{\infty} \sum_{l=0}^{\infty}\; {V_{k,l}(r,\theta)}^* \;  {V_{k,l}(s,\phi)}\;\,         
=\; \, \delta(r^2-s^2)\, \delta(\theta - \phi)
\end{split}
\end{align} 
and have the symmetries
\[
V_{l,k}(r, \theta) = {V_{k,l}(r, \theta)}^* = V_{k,l}(r, -\theta) . 
\]
 
Every function  $f(r,\theta) \in L^2(\D)$  can thus be written
\begin{equation}\label{2a}
f(r,\theta) = \sum_{k=0}^{\infty} \sum_{l=0}^{\infty}\; f_{k,l}\;\, V_{k,l}(r,\theta)
\end{equation}
where
\begin{equation}\label{3a}
f_{k,l} = \frac{1}{2\pi}\, \int_0^{2 \pi} d\theta \int_0^1 dr^2\; \, V_{k,l}(r,\theta)^*\;\; f(r,\theta)
\end{equation}
is the component along $V_{k,l}(r,\theta)$ of the function $f(r,\theta)$. In this paper 
we consider only normalized states, so that the Parseval identity gives for each state:

\begin{equation}\label{4a}
\frac{1}{2\pi}\, \int_0^{2 \pi} d\theta \int_0^1 dr^2\;  |f(r,\theta)|^2\,  =
\,  \sum_{k=0}^{\infty} \, \sum_{l=0}^{\infty} \, |f_{k,l}|^{~2} \;\;=\; 1.
\end{equation}

Restrictions on the values of $k$ and $l$ can be used as filters. $k+l >h \,(<h)$ 
is a high-pass (low-pass) filter in $r$, while $|k-l| >h\, (<h)$ is high-pass (low-pass)
filter in $\theta$ and a combination of the restrictions can be also considered. Of course,
using filters, the results must be multiplied for adequate factors
to obtain normalized states.

\section{ Operators and Algebra  in the disk space}

Now let us consider the differential applications :  
  \begin{align}\label{5a}
 \begin{split}
 {\cal A}_+ ~:=&~ \frac{e^{+{\bf i} \theta}}{2} \left[-(1-r^2)\frac{d}{dr} + r(k+l+2) +\frac{1}{r}(k-l)\right]
\sqrt{\frac{k+l+2}{k+l+1}}\\
 \vspace{.2cm}
 {\cal A}_- ~:=&~ \frac{e^{-{\bf i} \theta}}{2}  \left[+(1-r^2)\frac{d}{dr} + r(k+l)+\frac{1}{r}(k-l)\right] 
 \sqrt{\frac{k+l}{k+l+1}}\\
 \vspace{.2cm}
 {\cal B}_+ ~:=&~ \frac{e^{-{\bf i} \theta}}{2} \left[-(1-r^2)\frac{d}{dr} + r(k+l+2) -\frac{1}{r}(k-l)\right]
 \sqrt{\frac{k+l+2}{k+l+1}}\\
 \vspace{.2cm}
  {\cal B}_- ~:=&~ \frac{e^{+{\bf i} \theta}}{2} \left[+(1-r^2)\frac{d}{dr} + r(k+l) -\frac{1}{r}(k-l)\right]
  \sqrt{\frac{k+l}{k+l+1}}
  \end{split}
  \end{align} 
that, by inspection,  are the rising and lowering recurrence applications on the Zernike 
functions\, $V_{k,l}(r,\theta)$\; :
\begin{align}\label{8a}
\begin{split}
&{\cal A}_+\, V_{k,l}(r, \theta)\;=\; (k+1)\, ~V_{k+1,l}(r, \theta)\;, \qquad
{\cal A}_-\, V_{k,l}(r, \theta)\;=\; ~k\, ~V_{k-1,l}(r, \theta)\;, \\
&{\cal B}_+\,\; V_{k,l}(r, \theta)\;=\; (l+1)\, ~V_{k,l+1}(r, \theta)\;, \qquad
{\cal B}_-\, V_{k,l}(r, \theta)\;=\; ~l\, ~V_{k,l-1}(r, \theta)\; .
\end{split}
\end{align}

${\cal A}_\pm$ and ${\cal B}_\pm$ establish the recurrence relations but are not operators
because each application of   ${\cal A}_\pm$ or ${\cal B}_\pm$   modifies the parameters $k$ or $l$ to be read by the following applications. To obtain the true rising and lowering 
operators, we need indeed to introduce the operators\, $R$,\, $D_R$,\, $\Theta$, 
 $D_\Theta$, $K$, $L$:
\begin{align}\nonumber
\begin{split}
R~\, V_{k,l}(r,\theta)\,:=&\, ~  r\, V_{k,l}(r,\theta)~,\qquad D_R~\, V_{k,l}(r,\theta)\,:=\, \frac{d V_{k,l}(r,\theta)}{dr}~,~~~\\
\Theta~\, V_{k,l}(r,\theta)\,:=&\,  ~ \phi\, V_{k,l}(r,\theta)~,\qquad D_\Theta~\, V_{k,l}(r,\theta)\,:=\,  
\frac{d V_{k,l}(r,\theta)}{d\theta}~,~~~\\
K~\, V_{k,l}(r,\theta)\,:=&\; ~ k\, V_{k,l}(r,\theta)~,\qquad L~\, V_{k,l}(r,\theta)\;\;\;:=\; ~ l\, V_{k,l}(r,\theta)~;
\end{split}
\end{align}
and rewrite eqs.(\ref{5a}) as operators:
\begin{align}\label{7a}
\begin{split}
 A_+ :=&~~ \frac{e^{+ {\bf i} \Theta}}{2} \; \left[-(1-R^2) D_R + R (K+L+2) +\frac{1}{R}(K-L)\right]
 \sqrt{\frac{K+L+2}{K+L+1}}\; ,\\
 \vspace{1.4cm}
  A_- :=&~~ \frac{e^{ -{\bf i} \Theta}}{2} \; \left[+(1-R^2) D_R + R (K+L) +\frac{1}{R}(K-L)\right]~ \sqrt{\frac{K+L}{K+L+1}}\; ,\\
 \vspace{1.4cm}
 B_+:=&~~ \frac{ e^{- {\bf i} \Theta}}{2} \; \left[-(1-R^2) D_R + R (K+L+2) -\frac{1}{R}(K-L)\right]
 \sqrt{\frac{K+L+2}{K+L+1}}\; ,\\
 \vspace{1.4cm}
 B_-:=&~~ \frac{ e^{+ {\bf i} \Theta}}{2} \; \left[+(1-R^2) D_R + R (K+L) -\frac{1}{R}(K-L)\right]~\sqrt{\frac{K+L}{K+L+1}}\; .
 \end{split}
 \end{align}
Now, as\, $A_\pm$\, and\, $K$\, are operators, we can apply them iteratively on 
$V_{k,l}(r,\theta)$ and in particular we can calculate the action of the commutators: 
\[
[A_+,A_-]\; V_{k,l}(r,\theta)= -2(k+1/2)\; V_{k,l}(r,\theta)\,, \;\,\,\,\quad [K,A_\pm]\, V_{k,l}(r,\theta)= 
\pm \,V_{k \pm 1,l}(r,\theta)\,.
\]
So, defining\;\, $A_3 := K+1/2$\,,\; we find that\, $\{A_+,\, A_3,\, A_-\}$\, are on the\; 
$V_{k,l}(r,\theta)$ the generators of
one algebra\; $su(1,1)$:
\begin{equation}\label{alg1}
[A_+,\, A_-]\,=\,-2 A_3\;, \qquad [A_3,\, A_\pm]\,=\,\pm A_\pm \,.
\end{equation}
Analogously
\[
[B_+,B_-]\; V_{k,l}(r,\theta)= -2(l+1/2)\; V_{k,l}(r,\theta) \,,\,\,\quad [L,B_\pm] V_{k,l}(r,\theta)=
 \pm \,V_{k,l \pm 1}(r,\theta)
\]
exhibit that\, $\{B_+,\; B_3:= L+1/2\; ,\, B_-\}$\, are on the\; $V_{k,l}(r,\theta)$
the generators of another Lie
algebra\; $su(1,1)$:
\begin{equation}\label{alg2}
[B_+,\, B_-]\,=\,-2 B_3 \qquad [B_3,\, B_\pm]\,=\,\pm B_\pm \,.
\end{equation}
Finally, as $A_i$ and $B_j$ commute on the\; $V_{k,l}(r,\theta)$, the algebra is completed by 
\begin{equation}\label{alg3}
[A_i, B_j] = 0\, .
\end{equation}
Thus in the vector space of images on the disk, 
eqs.(\ref{alg1} -\ref{alg3}) define a differential realization
of the 6 dimensional Lie algebra\; $su(1,1)\oplus su(1,1)$ .

We can now calculate on \;$V_{k,l}(r,\theta)$\, the Casimir invariants of the two $su(1,1)$:
\[
C_A\;  V_{k,l}(r,\theta)\,=\,\left[\frac{1}{2} \{ A_+, A_- \} - A_3^2\right]\, V_{k,l}(r,\theta)\, =\, \frac{1}{4}\, 
V_{k,l}(r,\theta) ,
\]
\[
C_B\;  V_{k,l}(r,\theta)\,=\,\left[\frac{1}{2} \{ B_+, B_- \} - B_3^2\right]\, V_{k,l}(r,\theta)\, =\, \frac{1}{4}\, 
V_{k,l}(r,\theta) .
\]
As the Casimir of the discrete series\, $D^+_j$  of\, $su(1,1)$\, is\; $j(1-j)$\cite{Ba}\; with 
$j=1/2, 1,..$, the space $\{V_{k,l}(r,\theta)\}$\, is isomorphic to $\{|k,l\rangle |k,l=0,1,2\dots\}$,
the unitary irreducible representation\; $D^+_{1/2} \otimes D^+_{1/2}$ 
of the group\; $SU(1,1)\otimes SU(1,1)$, where indeed the action of the generators 
of the algebra is:
\begin{align}\label{8a}
\begin{split}
&A_+\, |k,l\rangle\;=\; (k+1)\, ~|k+1,l\rangle\;, \qquad
\;B_+\, |k,l\rangle\;\;=\; (l+1)\, ~|k,l+1\rangle\;, \quad\\
&A_3\,\; |k,l\rangle \,=\,~~ k\,~ |k,l\rangle \;,\;\;\;\qquad\quad\;\, 
~~~~~B_3\,\; |k,l\rangle \,\;=\,~~ l\,~ |k,l\rangle \;,\;\;\;\quad \\
&A_-\, |k,l\rangle\;=\; ~~k\, ~|k-1,l\rangle\;, \qquad
~~~~~~B_-\, |k,l\rangle\;\;=\; ~\;l\, ~|k,l-1\rangle\; .
\end{split}
\end{align}

\indent Now we move from the generators of the algebra\; $su(1,1)\otimes su(1,1)$ to
their products that belong to the associated universal 
enveloping algebra (UEA). It is the algebra\;   
UEA[$su(1,1)\oplus su(1,1)$] 
constructed on the ordered monomials\;$A_+^{\alpha_1} A_3^{\alpha_2} A_-^{\alpha_3} B_+^{\beta_1} B_3^{\beta_2} B_-^{\beta_3}$ 
(where\; $\alpha_i$ and ${ \beta_j}$\,  are natural
numbers) submitted to the relations (\ref{alg1}-\ref{alg3}). 
So every operator ${\cal O} \in$ UEA[$su(1,1)\oplus su(1,1)$]\; can be written as
\begin{equation}\label{13a}
{\cal O} = \sum_{\bar{\alpha}, \bar{\beta}} {\cal O}_{\bar{\alpha}, \bar{\beta}} = 
\sum_{\bar{\alpha}, \bar{\beta}} c_{\bar{\alpha}, \bar{\beta}} \; A_+^{\alpha_1} A_3^{\alpha_2} 
A_-^{\alpha_3} B_+^{\beta_1} B_3^{\beta_2} B_-^{\beta_3} .
\end{equation}
where\, $c_{\bar{\alpha}, \bar{\beta}}$ are arbitrary complex functions of           
${\bar \alpha} = (\alpha_1, \alpha_2, \alpha_3)$ and 
${\bar \beta} = (\beta_1, \beta_2, \beta_3)$.
Since\, $\{V_{k,l}(r,\theta)\}$\, is a differential representation of the algebra\; 
$su(1,1)\oplus su(1,1)$,
it is also a differential representation of the $UEA[su(1,1)\oplus su(1,1)]$.

Because the representation $D^+_{1/2} \otimes D^+_{1/2}$ is unitary and 
irreducible the set
of unitary operators acting on the space $L^2(\D)$, $\{{\cal O}[L^2(\D)]$, is isomorphic 
to the set of operators acting on $D^+_{1/2} \otimes D^+_{1/2}$.
Therefore all invertible operators that transform disk images into disk images
can be written in the form (\ref{13a}) and belong to the
$UEA[su(1,1)\oplus su(1,1)]$.

In concrete:  all transformations of an arbitrary image in whatsoever other image 
can be realized by means of iterated applications of the operators (\ref{7a}). 

\section{Applications to image processing}

In physics a fundamental point of image processing is 
that every image is the result of a measure and each measure has a
measure error.
This implies that all numbers in the preceding sections --that in
mathematics are unlimited-- in physics can be considered always finite, because 
the limited level of accuracy.
All problems related to the rigged Hilbert space are thus irrelevant as, in finite dimensions,
rigged Hilbert spaces and Hilbert spaces are equivalent 
(see \cite{Ce17}, \cite{CeGaOl16} and \cite{CeGaOl17}).

Let us start depicting the procedure that, starting from one image\, $|f(r,\theta)|^2$\,, by
means of an operator\, ${\cal O}$\, of the\, $UEA$, gives us another image\, $|g(r,\theta)|^2$\,.

As each image\, $|f(r,\theta)|^2$\, does not depend from the phases, it is completely  
determined by $|f(r,\theta)|$. Thus we look for the
components along $V_{k,l}(r,\theta)$ of $|f(r,\theta)|$:
\begin{equation}\label{11.5}
f_{k,l} = \frac{1}{2\pi}\, \int_0^{2 \pi} d\theta \int_0^1 dr^2\;\, V_{k,l}(r,\theta)^*\;\; |f(r,\theta)|
\end{equation}
where, as $|f(r,\theta)|$ is real,  $f_{l,k} = (f_{k,l})^*$.

Because of the measure errors, the relevant values of $k$ and $l$ are limited to the $k_M$ and $l_M$ such that the Parseval identity eq.(\ref{4a}) is satisfied in the approximation appropriate  to the accuracy
of the experimental result\, $|f(r,\theta)|^2$. We thus have:
\begin{equation}\label{12a}
|f(r,\theta)|\; \approx\; \sum_{k=0}^{k_M} \sum_{l=0}^{l_M}\;  f_{k,l}\;\, V_{k,l}(r,\theta) .
\end{equation}
Let us describe now how a transformation ${\cal O}$ allows to obtain from the image  
$|f(r,\theta)|^2$ a new image $|g(r,\theta)|^2$.

The operator ${\cal O}$ will be
\begin{equation}\label{33}
{\cal O} =\; \sum_{\bar{\alpha}, \bar{\beta}} {\cal O}_{\bar{\alpha}, \bar{\beta}} \;=\; 
\sum_{\bar{\alpha}, \bar{\beta}}\; c_{\bar{\alpha}, \bar{\beta}}\;\; A_+^{\alpha_1} A_3^{\alpha_2} 
A_-^{\alpha_3} B_+^{\beta_1} B_3^{\beta_2} B_-^{\beta_3} .
\end{equation}
where in physics the sums on $\bar{\alpha}$ and $\bar{\beta}$ can be assumed to be finite.

\noindent We thus write
\[
{\cal O}_{\bar{\alpha}, \bar{\beta}}\;\, |f(r,\theta)|\; =\;  \sum_{k=0}^{k_M} \sum_{l=0}^{l_M} 
\; f_{kl}\, \; c_{\bar{\alpha}, \bar{\beta}}\;\; A_+^{\alpha_1} A_3^{\alpha_2} 
A_-^{\alpha_3} B_+^{\beta_1} B_3^{\beta_2} B_-^{\beta_3}\;\; V_{k,l}(r,\theta)\,\,,
\]
then --by means of iterated applications of operators (\ref{7a})--
we calculate
\[
A_+^{\alpha_1} A_3^{\alpha_2} 
A_-^{\alpha_3} B_+^{\beta_1} B_3^{\beta_2} B_-^{\beta_3}\; V_{k,l}(r,\theta)
\]
finding the coefficients\, $g_{k,l}$ that satisfy
\[
A_+^{\alpha_1} A_3^{\alpha_2} 
A_-^{\alpha_3} B_+^{\beta_1} B_3^{\beta_2} B_-^{\beta_3}\; V_{k,l}(r,\theta)\; =\;
g_{k+\alpha_1-\alpha_3, l+\beta_1-\beta_3}\; V_{k+\alpha_1-\alpha3,\, l+\beta_1-\beta_3}(r,\theta) .
\]
Thus we have
\[
{\cal O}_{\bar{\alpha}, \bar{\beta}}\;\, |f(r,\theta)|\, =\,  \sum_{k,l=0,0}^{k_M,l_M}\; f_{kl} \;\,
 c_{\bar{\alpha}, \bar{\beta}}\;\; g_{k+\alpha_1-\alpha_3,\, l+\beta_1-\beta_3}\; 
 V_{k+\alpha_1-\alpha_3,\, l+\beta_1-\beta_3}(r,\theta)\; ,
\]
that, combined with eq.(\ref{33}),  allows to obtain:
\[
 g(r,\theta)\; = \;{\cal O}\; |f(r,\theta)| 
\]
from which  $|g(r,\theta)|^2$, the transformed image under\, ${\cal O}$ of\, $|f(r,\theta)|^2$,
is obtained. 

Analogous procedure can be applied to obtain the operator ${\cal O}$ from 
 $|g(r,\theta)|$ and $|f(r,\theta)|$. 

To conclude, let us consider now a possible application to improve, by means of AIP, 
an image obtained by a flawed instrument.
We start observing a null signal that, with a perfect tool, would give\, $|f(r,\theta)|
= |V_{0,0}(r,\theta)|$.
A defective instrument, on the contrary, will give a perturbed image\, $|f(r,\theta)|^2$ such that
\begin{equation}\label{15}
|f(r,\theta)| \,=\,  \sum_{k=0}^{k_M} \sum_{l=0}^{l_M}\;  f_{k,l}\; V_{k,l}(r,\theta)  \,=\, 
 \sum^{k_M}_{k=0} \sum^{l_M}_{l=0}\;  f_{k,l}\; 
\frac{A_+^k\; B_+^l}{k! \;l!}\;  V_{0,0}(r,\theta)
\end{equation}
where $f_{k,l} \;(= f_{l,k}^*)$ are the parameters, obtained from eq.({\ref{11.5}), that characterize the
distortion of the null image made by the instrument.

So that the operator that eliminates  the defects 
of the instrument is
\[
\left(\sum^{k_M}_{k=0} \sum^{l_M}_{l=0}  f_{k,l}\; 
\frac{A_+^k\, B_+^l}{k!\; l!}\right)^{-1}\, .
\]
 If the observation of the object we are interested in gives, with $g_{k,l} \;= (g_{l,k})^*$, 
\begin{equation}\label{15}
|g(r,\theta)| \;=\;  \sum_{k=0}^{k_{M'}} \sum_{l=0}^{l_{M'}}\,  g_{k,l}\; V_{k,l}(r,\theta)  \;= \;
 \sum^{k_{M'}}_{k=0} \sum^{l_{M'}}_{l=0}\; g_{k,l}\; 
\frac{A_+^k\; B_+^l}{k! \;l!}\;  V_{0,0}(r,\theta)
\end{equation}
\noindent the final cleaned image  will be
\begin{equation}
\left|
 \left(
\sum^{k_{M'}}_{k=0} \sum^{l_{M'}}_{l=0}\; g_{k,l}\;
\frac{A_+^k\; B_+^l}{k! l!}\;
\right)
\left(
\sum^{k_M}_{k=0} \sum^{l_M}_{l=0}\, f_{k,l}\; 
\frac{A_+^k\; B_+^l}{k! l!} 
\right)^{-1} 
V_{0,0}(r,\theta)\;
\right|^2   ,
\end{equation}
formula that can be easily calculated in series
as all operators commute.



\end{document}